\documentstyle[twoside,fleqn,epsfig,latexsym,amssymb,espcrc3,graphicx,graphics]{article}

\newcommand{\ra}{\rightarrow}

\newcommand{\be}{\begin{equation}}
\newcommand{\eea}{\end{eqnarray}}
\newcommand{\beq}{\begin{equation}}
\newcommand{\eeq}{\end{equation}}
\newcommand{\bea}{\begin{eqnarray}}
\newcommand{\ee}{\end{equation}}

\newcommand{\gev}{{\rm GeV}}

\newcommand{\nn}{\nonumber}

\newcommand{\<}{\langle}
\renewcommand{\>}{\rangle}

\newcommand{\vettx}{\mbox{\bf{x}}}
\newcommand{\vetty}{\mbox{\bf{y}}}
\newcommand{\vettq}{\mbox{\bf{q}}}

\title{\vspace*{-0.1cm}Extraction of $K\ra\pi\pi$ Matrix Elements with Wilson Fermions\thanks{Talk presented by Mauro Papinutto}}

\author{
$\mathrm{SPQ_{CD}R}$ Collaboration:~Ph.~Boucaud\address{\vspace*{-0.28cm}Laboratoire de Physique Th\'eorique (b\^at.210),
Universit\'e de Paris XI, 91405 Orsay Cedex, France}, 
V.~Gim\'enez\address{\vspace*{-0.28cm}Dep.~de F\'{\i}s.Te\`orica and IFIC, Univ.~de Val\`encia, Dr.~Moliner 50, E-46100, Burjassot, Val\`encia,
Spain}, 
C.-J.~D.~Lin\address{\vspace*{-0.28cm}Department of Physics and Astronomy, University of
Southampton, Southampton SO17 1BJ, England},
V.~Lubicz\address{\vspace*{-0.28cm}Dip. di Fisica, Univ. di Roma Tre and INFN - Roma III, 
Via della V. Navale 84, I-00146 Rome, Italy},   
G.~Martinelli\address{\vspace*{-0.28cm}Dip. di Fisica, 
Universit\`a ``La Sapienza" and INFN-Roma I, P.le A. Moro 2, I-00185 Rome,
Italy},
M.~Papinutto\address{\vspace*{-0.0cm}Dip. di Fisica, Universit\`a di Pisa and INFN - Pisa, 
Via Buonarroti 2, I-56100 Pisa, Italy}
F.~Rapuano$^{\rm e}$,
C.~T.~Sachrajda$^{\rm c}$}
\begin{document}
\newcommand{\sze}{\small}

\begin{abstract}
\vspace*{-0.1cm}
We present the status of a lattice calculation for the $K\ra\pi\pi$ 
matrix elements of the $\Delta S\!\!=\!\!1$ effective weak Hamiltonian, 
directly with two pion in the final state. 
We study the energy shift of two pion in a finite
volume both in the $I\!\!=\!\!0$ and $I\!\!=\!\!2$ channels. 
We explain a method to avoid the Goldstone pole contamination in  
the computation of renormalization constants 
for $\Delta I\!\!=\!\!3/2$ operators. Finally we show some preliminary 
results for the matrix elements of $\Delta I\!\!=\!\!1/2$ operators.
Our quenched simulation is done at $\beta\!\!=\!\!6.0$, with Wilson fermions, 
on a $24^3\times64$ lattice.
\vspace*{-0.6cm}
\end{abstract}
 
\maketitle

\section{General Strategy}
\vspace{-0.1cm} 
Kaon weak decay amplitudes can be described 
in terms of matrix elements (ME's) of the $\Delta S\!\!=\!\!1$ effective 
weak Hamiltonian. ${\cal H}^{\Delta S=1}$ is written as a linear 
combination of a complete basis of 
renormalized local operators (OP's) $Q_{i}(\mu)$, where $\mu$ is the
renormalization scale. The most relevant
contributions in the computation of the weak amplitudes 
${\cal A}_{I=0,2}$ and of $\epsilon'/\epsilon$ are given by the 
$K\!\!\ra\!\!\pi\pi$
ME's of $\hat Q^{+}$, $\hat Q^{-}$, $\hat Q_6$, $\hat Q_7$
and $\hat Q_8$ (see below and Ref.~\cite{lin,damir}). 
In order to compute these ME's from lattice QCD, 
one has to renormalize the bare (divergent) lattice OP's.
Two methods are possible on the lattice:\\
$1)$ Compute $\<0|\hat Q_{i}(\mu)|K\>$ and $\<\pi|\hat
Q_{i}(\mu)|K\>$ and then derive $\<\pi\pi|\hat Q_{i}(\mu)|K\>_{I=0,2}$  
using soft pion theorems~\cite{claude}. In this case, besides the
difficulties in controlling chiral behaviour, a major problem is
the inclusion of higher order contributions in the chiral expansion
which should reconstruct the final state interaction (FSI). In fact, in
this approach, only $K\ra\pi\pi$ ME's at lowest order in ChPT 
can be obtained (see Refs.~\cite{kpi}).\\
\mbox{$2)$\hspace{-0.05cm} Compute directly\hspace{-0.05cm} $\<\pi\pi|\hat
Q_{i}(\mu)|K\>_{I=0,2}$.\hspace{-0.1cm} The main}
difficulty in this case is the relation between ME's in a (Euclidean) finite  
volume and the corresponding physical infinite volume
ones. Even if, in principle, it is possible to take 
into account FSI exactly~\cite{LL,LMST}, in practice these methods are
numerically very demanding. So, with present computing power, only
simulation at unphysical kinematics~\cite{lin} may be achieved, and ChPT
is still needed to extrapolate to the physical point.  
\par In this work we choose the second strategy.
To extract ME's we place the local OP $Q_i$ in the origin and we use
three local interpolating fields (e.g. the pseudoscalar density
$P^{ab}=\bar \psi^a \gamma_5 \psi^b$), 
one at time $t_2=10$ which annihilates one $\pi$ in the two
pion state and one at time $t_K=54\equiv-10$ which creates 
a $K$ (both with momentum zero). As explained below $t_2$ ($t_K$) must 
be chosen large enough in modulus for the two pion state (the kaon 
state) to be already asymptotic. The third one annihilates at 
time $t_1$ (not fixed) the second $\pi$ with momentum either $0$ or $2\pi/L$.
As shown in~\cite{LMST}, in the limit $T/2\! >\! t_1\!\gg\!
t_2\!\gg0$,\hspace{0.1cm} $T\!\gg\!t_K\!>\!T/2$ we have

\vspace*{-0.4cm}
\bea
\hspace{0.2cm}\frac{\<0|T[\hat \pi_1(t_1)\hat \pi_2(t_2) Q_i(0)
\hat K(t_K)]|0\>_V}{G_\pi(t_1)G_\pi(t_2)G_K(t_K)}Z_\pi^2 Z_K
\,\longrightarrow\nn
\eea
\vspace*{-0.45cm}
\bea
\longrightarrow |\<\pi\pi|Q_i|K\>| \cos\delta(W)\;\textrm{\large e}^{^{-(W-2M_\pi)t_2}}\!\!+O\Big(\frac{1}{L}\Big)\qquad\nn
\eea 
\vspace*{-0.38cm}

\noindent where $G_\pi(t)\!\!=\!\!\<0|T[\hat \pi(t)\hat \pi^\dag(0)]|0\>_V$, $Z_\pi\!\!=\!\!|\<0|\hat \pi|\pi\>|$ (the definition of $G_K(t)$ and $Z_K$ is analougous), 
$|\pi\pi\>$ is the lower two pion state with the same 
momentum of $\hat \pi_1(t_1)$.
$\Delta E=W-2M_\pi$ is the energy shift of two pions in a finite volume
(FV) while $\delta(W)$ is the strong interaction phase (which depends on
the isospin and on the energy of the two pion state). In order 
to extract the physical ME, one needs to compute FV
corrections, which are represented by the correction to the ME
(indicated above by the term $O(1/L)$) and by the FV energy shift. 
We focus now on this last issue. Theoretical
predictions~\cite{luscher,bergolt} show that the 
factor $\exp(-\Delta E t_2)$ should give relevant corrections: 
of order $5\div10\%$ in the $\Delta I=3/2$
channel and of order $25\%$ (or larger) in the $\Delta I=1/2$ one.  
\begin{figure}[!t]
\vspace*{-0.5cm}
\mbox{
\hspace*{-0.6cm}
\epsfig{figure=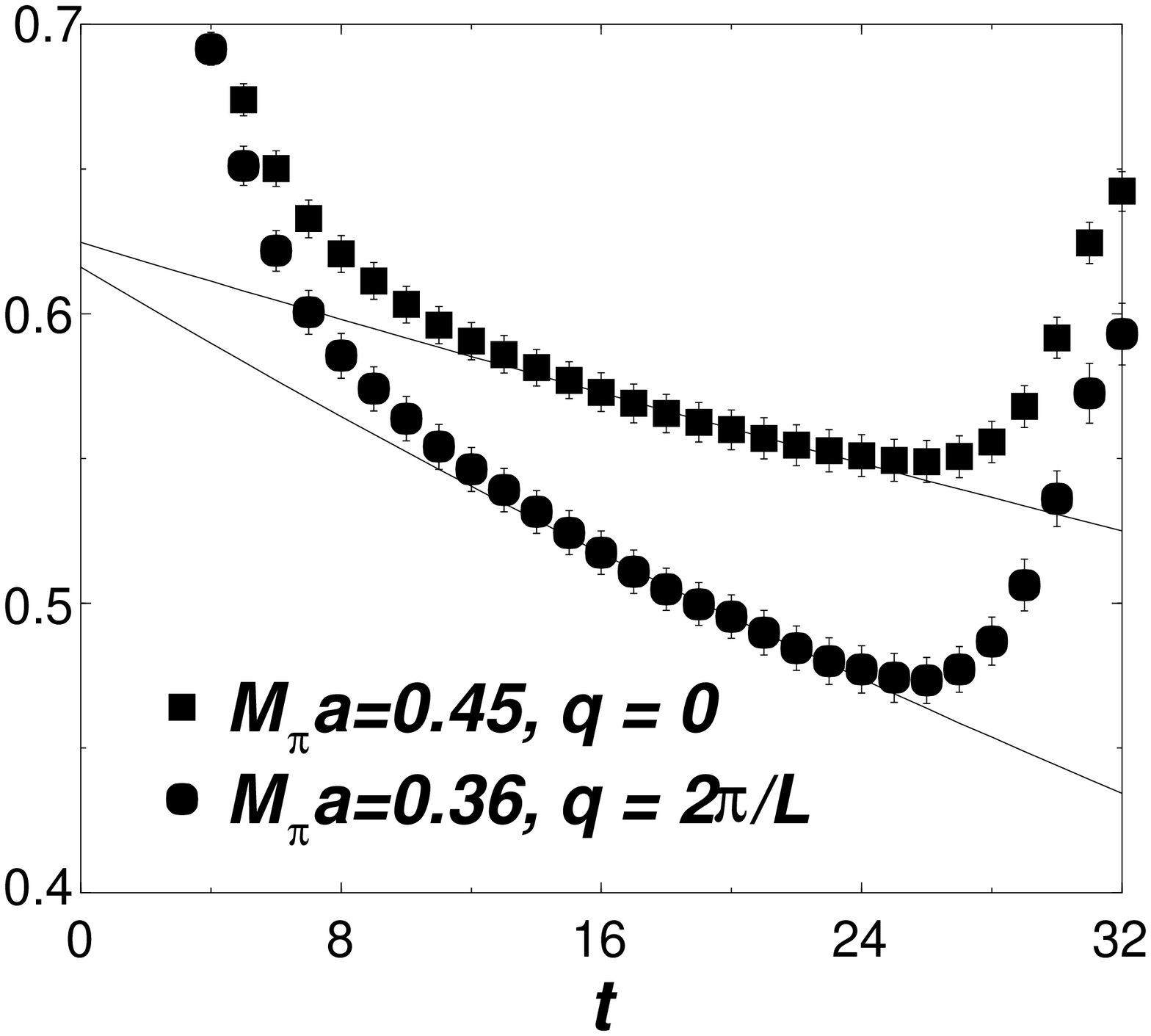,angle=0,width=0.57\linewidth}
\put(3,115){\epsfig{figure=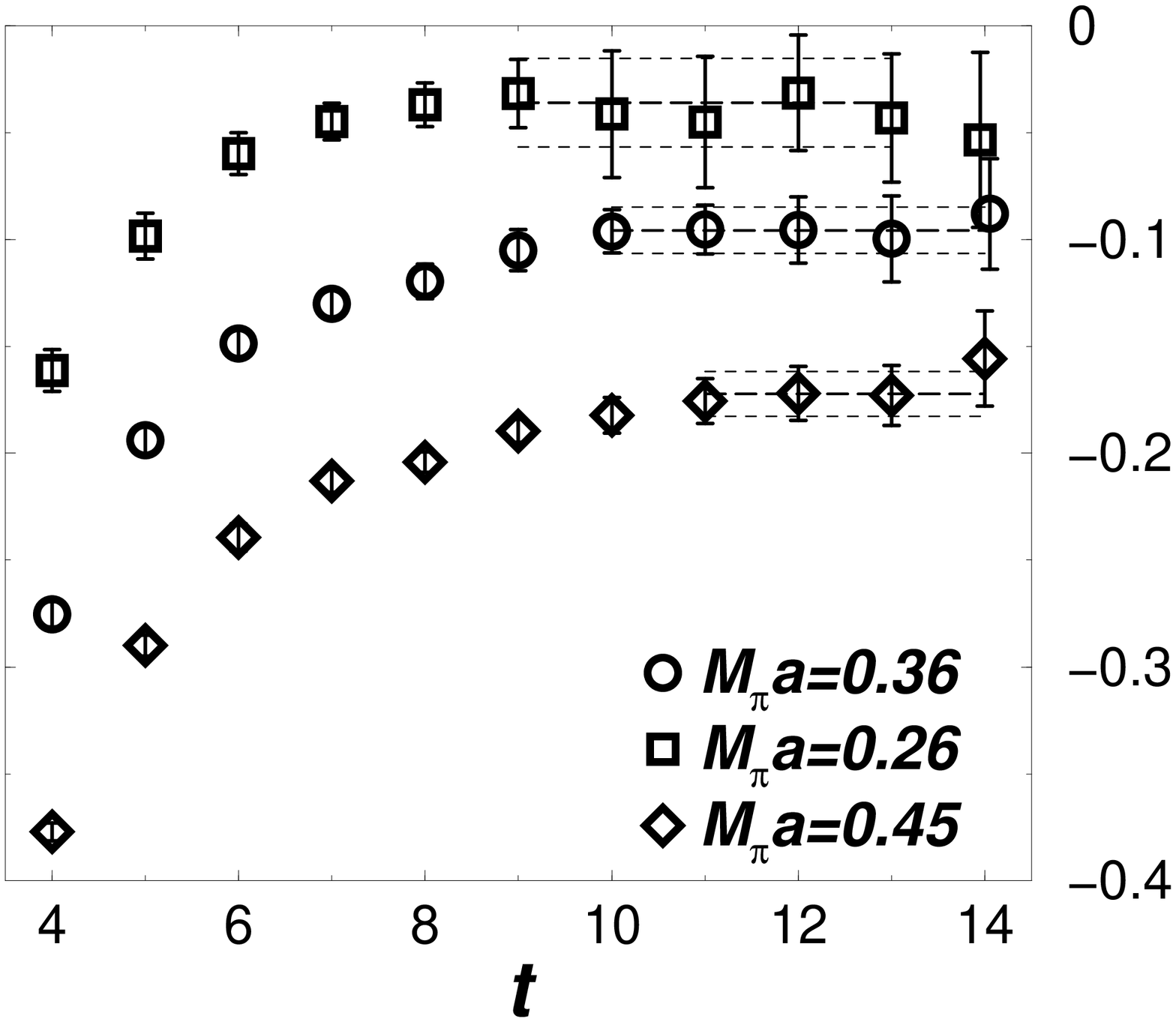,angle=-90,width=0.575\linewidth}}}
\put(-19,10){\tiny $\frac{T}{2}\!\!\!=$}
\vspace{-1.2cm}
\caption{\sl \small a. $G_{4\pi}(t)/G_\pi(t)^2$\hspace{0.2cm} \newline
\hspace*{1.5cm}b. effective mass of\hspace*{0.2cm}$G_{2\pi S}(t)/G_\pi(t)^2$}
\label{eshift}
\vspace*{-0.7cm}
\end{figure}     
\vspace*{-0.2cm}    
\section{Finite Volume Energy Shift} 
\vspace*{-0.2cm}    
To extract $\Delta E$ we study the four point function 
$G_{4\pi}(t)=\<0|T[\hat \pi_4(t)\hat \pi_3(t)\hat \pi^\dag_2(0)\hat
\pi^\dag_1(0)]|0\>_V$. Since the computation of the $I=0$ correlator 
require the computation of $T$ $S_1$ propagators, we use
$G_{4\pi}(t)$ only for $I=2$. We have $\hat \pi_2(0)\hat
\pi_1(0)=P(0)P(0)$ while $\hat \pi_4(t)\hat
\pi_3(t)=\sum_{\vettx}P(\vettx,t) \exp{i\vettq\vettx}
\sum_{\vetty}P(\vetty,t)$ with $\vettq\!=\!0,2\pi/L$ and 
with the appropriate flavour structure.
Due to the possibility of propagation both of a two pion state in one temporal
direction and of a state composed by two single pion states in opposite
directions, in the region $T\gg t\gg 0$ one has
\vspace*{-0.3cm}
\bea
G_{4\pi}(t)/G_\pi(t)^2\longrightarrow\frac{C_{\pi\pi}
\cosh(W(t-\frac{T}{2}))+C_\pi}{\cosh(M_\pi(t-\frac{T}{2}))^2}\nn
\eea
\vspace*{-0.2cm}

\noindent where $C_{\pi\pi}$ and $C_{\pi}$ depend on the ME's of the interpolating
fields and on the energy shift. We thus make a fit with $A \exp(-\Delta E t)$
in the region $T/2\gg t \gg0$ (see Fig.~\ref{eshift}.a).
\par In the $I=0$ channel we use instead $G_{2\pi S}(t)=
\<0|T[S(t)\hat \pi^\dag_2(0)\hat\pi^\dag_1(0)]|0\>_V$ where $S$ is the scalar
density summed over the space and   
$\hat \pi_2(0)\hat \pi_1(0)=\Pi(0)\sum_{\vettx}P(\vettx,0)$, with $\Pi$
equal either to $P$ or to $A_0$ (the temporal component
of the axial current). We have computed only the connected part, 
because we expect the disconnected one to be
negligible since it is suppressed by the OZI rule. To check this
assumption, a computation of the disconnected part is presently under way. 
We observe empirically that, for $\Pi\!\!=\!\!A_0$,
the noise is smaller and the plateaux start at smaller times
(already $t\!\simeq\!10$ seems sufficients, see Fig.~\ref{eshift}.b).  
Results for both $I\!\!=\!\!2$ and $I\!\!=\!\!0$ channel are shown in 
Table~\ref{deltae}. We compare numerical results both with L\"uscher
formula~\cite{luscher}
\vspace*{-0.1cm}
\bea
\Delta E_L\!=\!-\!\frac{4\pi a^{I=0,2}_0}{M_\pi L^3}\!+\!{\cal
O}\Big(\frac{1}{L^4}\Big)\;\textrm{with}\;
\left\{ \begin{array}{c}
\!a^{I=2}_0\!=\!-\frac{M_\pi}{8\pi F_\pi^2}\\
\!a^{I=0}_0\!=\!\frac{7 M_\pi}{16\pi F_\pi^2}\end{array}\right.\nn
\eea
\vspace*{-0.2cm}

\noindent (where $a^{I=0,2}_0$ is the tree level result in ChPT)
and with its version $\Delta E_{BG}$ in 1-loop quenched ChPT~\cite{bergolt}
which, in the $I\!=\!0$ case, shows enhanced finite volume
correction. For $I\!\!=\!\!2$ there is very good agreement 
between numerical values and theoretical predictions. 
For $I\!\!=\!\!0$ there is a striking disagreement 
for the two heaviest masses while in the lightest case, also due to the large
errors, they are compatible. Moreover the dependence with the mass of
the pion is different to the one predicted. This could be due to the
presence, for large unphysical quark masses, of a
stable scalar particle below the two pion state. In this case
the behaviour of the shift measured should have only exponentially small
correction in the volume. Instead the ``genuine'' FV energy shift of the
two pion state has a power dependence on the volume. To clarify the
situation we are thus currently varying the volume of our simulations 
and trying to reach smaller masses.        

\small{\begin{table}[!t]
\vspace*{-0.5cm}
\hspace*{0.1cm}
\begin{tabular}{cccc}
\hline\hline
$M_\pi$ & 0.446(2) & 0.361(2) & 0.260(2)\\
\hline
\multicolumn{4}{c}{$I=2\qquad 750\; \textrm{conf.}$} \\ 
\hline$\Delta E^N_{\vettq=0}$& 5.4(6)(5) & 6.2(7)(5) & 6.6(9)(5)\\
$\Delta E^N_{\vettq=\frac{2\pi}{L}}$& 8.6(8)(5) & 10.9(11)(8) & 15.2(27)(8)\\
$\Delta E_L$& 4.9(2) & 5.6(2) & 6.8(3)\\
$\Delta E_{BG}$& 5.3(2) & 6.1(2) & 7.3(3)\\                           
\hline
\multicolumn{4}{c}{$I=0\qquad 600\; \textrm{conf.}$} \\ 
\hline
$\Delta E^N$& -172(10) & -96(11) & -36(21)\\
$\Delta E_L$& -17.1(5) & -19.7(7) & -23.7(11)\\
$\Delta E_{BG}$& -15.4(4) & -18.0(6) & -22.0(9)\\
\hline\hline                                 
\end{tabular}
\vspace*{0.2cm}
\caption{\sl \small Results in lattice units multiplied by $\!10^{3}\!$}
\label{deltae}
\vspace*{-0.7cm}
\end{table}}
\normalsize{
\vspace*{-0.2cm}    
\section{Renormalization and Goldstone Pole}
\vspace*{-0.2cm}
The renormalization  of $\Delta I=3/2$ 4-fermion OP's, 
which mix only with OP's of the same dimension, may be
achieved with the method of ref.~\cite{npren}.
This requires the existence of a window $\Lambda_{QCD}\ll \mu \ll 1/a$, where
the first inequality serves to avoid pure non-perturbative effects like the
coupling to the Goldstone boson. In some 
cases~\cite{alain,dawson} this makes the chiral limit ill-defined and
causes a systematic error in the extrapolation. It is easy to show that 
for parity conservig OP's this coupling gives both a single and a 
double pole, while in
the parity violating (PV) case only a single pole is present. For the sake of
illustration, we explain the procedure to remove this unwanted
contribution only in the PV case. Consider the amputated Fourier
transform (at equal external momenta $p$) 
of the Green functions $G_i=\<\psi(x_1)\bar\psi(x_2)
O_i(0) \psi(x_3)\bar\psi(x_4)\>$. By projecting this
array onto the possible indipendent Dirac structure we obtain a 
$5\times5$ matrix
$D(M^2_\pi,p^2)$ from which the matrix of RC's is obtained~\cite{bibbia} 
by $Z(\mu a)=Z_\psi^2 (D^T)^{-1}|_{p^2=\mu^2}$, where $Z_\psi$ 
is the quark field RC. The pole is present e.g. in
$D_{33}$, which corresponds to the OP $O_3=SP-PS$. This is
evident from Fig.~\ref{goldpole}.a. To subtract it, it is possible 
either to fit each element of $D$ directly with 
$A(p^2) + B(p^2)/M^2_\pi + C(p^2) M^2_\pi$ or to construct the combination 
\begin{figure}[!t]
\vspace*{-0.4cm}
\mbox{
\hspace*{-0.6cm}
\epsfig{figure=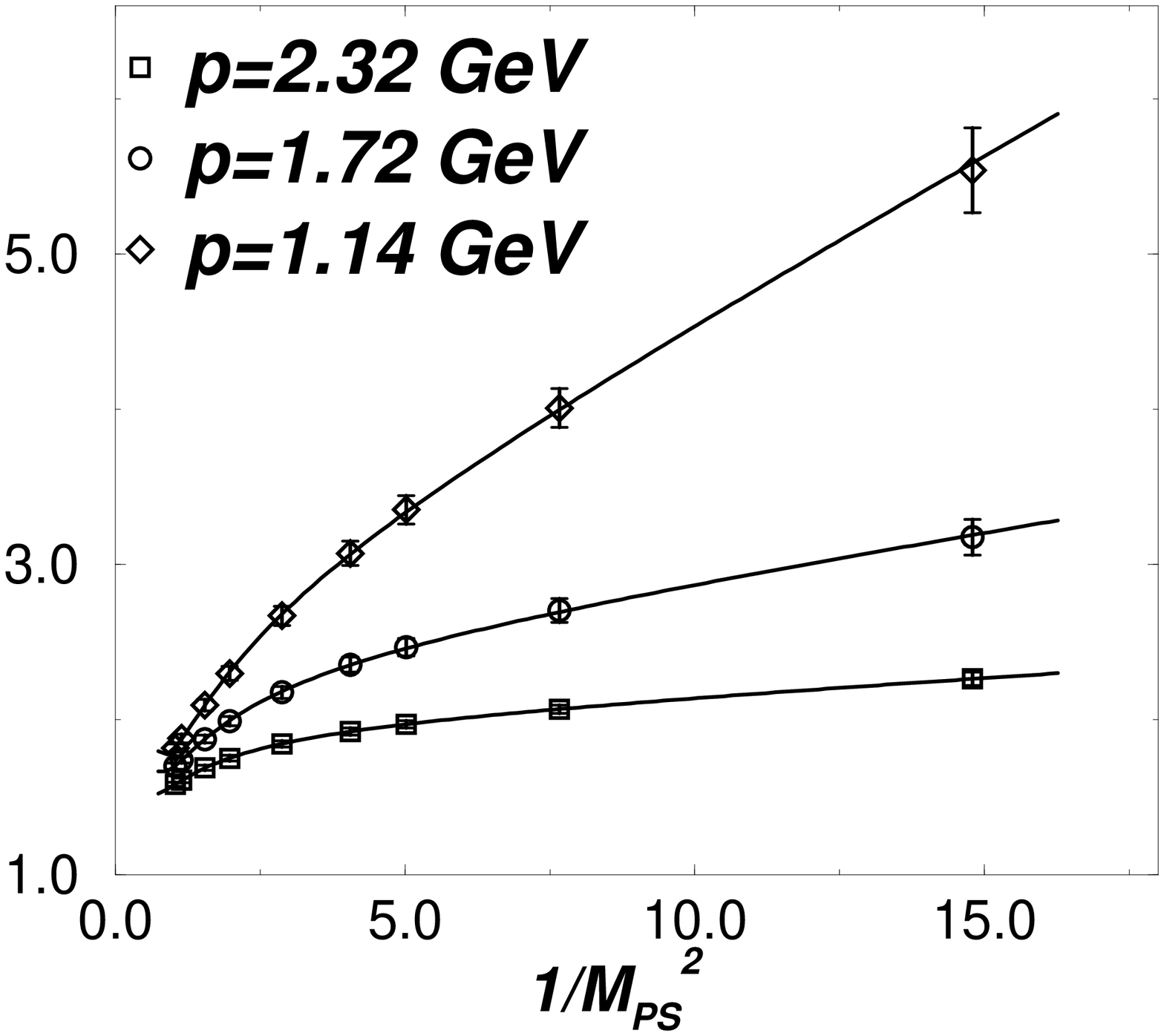,angle=-90,width=0.61\linewidth}
\put(-4,-1){\epsfig{figure=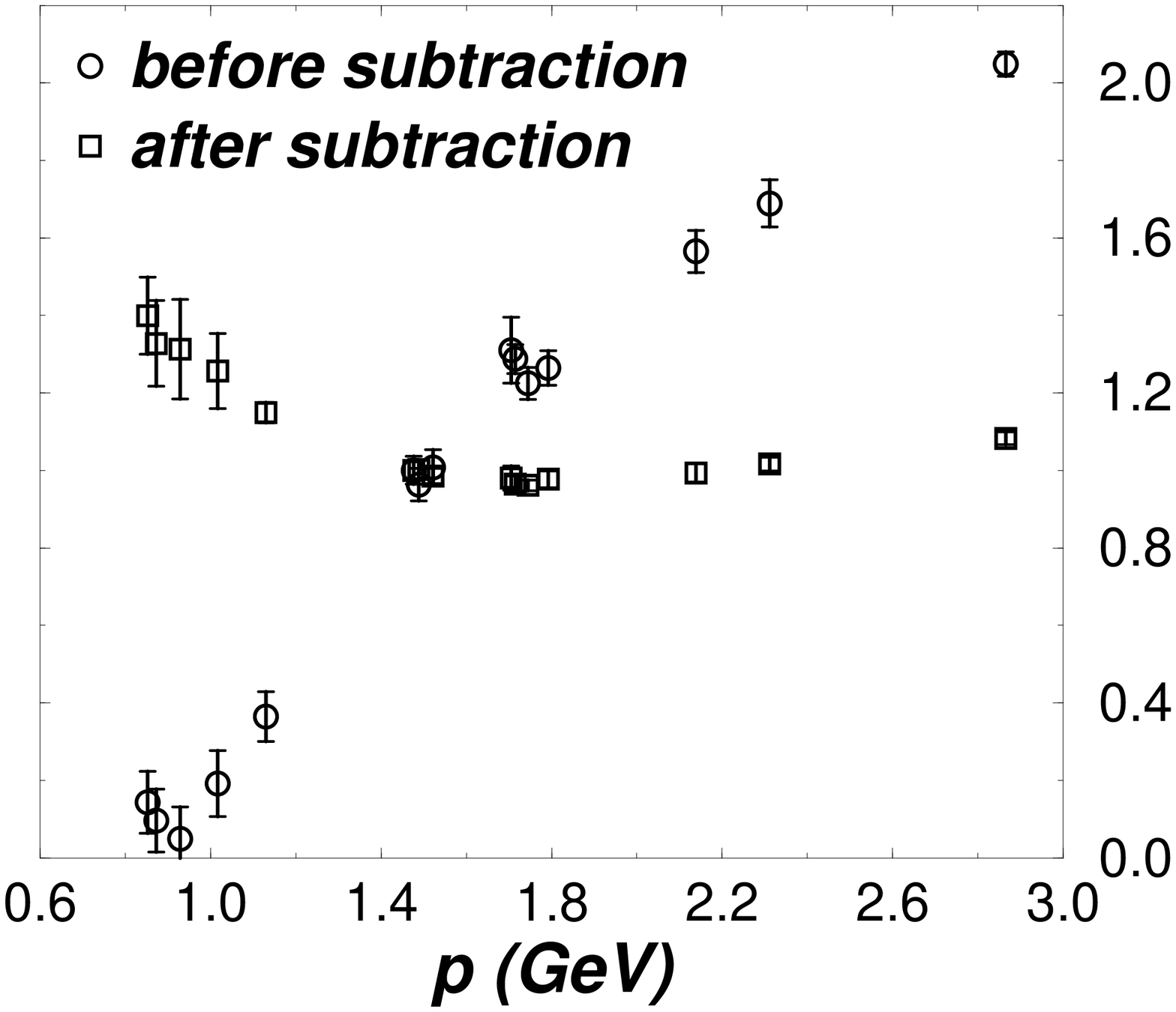,angle=-90,width=0.64\linewidth}}}
\vspace{-1.1cm}
\caption{\sl \small a. $D_{33}(M_\pi^2,p^2)$ b.
$(Z(\mu) Z^{^{-1}}\hspace*{-0.35cm}(\mu_0)U^{^{-1}}\hspace*{-0.35cm}(\mu,\mu_0))_{_{33}}$}
\label{goldpole}
\vspace*{-0.8cm}
\end{figure}     
\vspace*{-0.0cm}
\be
\qquad\frac{D(M_1^2,p^2)M_1^2 - D(M_2^2,p^2)M_2^2}{M_1^2-M_2^2}
\ee
\vspace*{-0.1cm}
(for non-equal masses), which automatically
cancels the pole~\cite{giusti}, and fit it with
$\tilde A(p^2) + \tilde C(p^2) (M^2_1+M^2_2)$. Finally, by using either
$A(p^2)$ or $\tilde A(p^2)$, we obtain two determination of the 
RC's perfectly compatible with each other. In Fig.~\ref{goldpole}.b
the behaviour of $(Z(\mu) Z^{^{-1}}\hspace*{-0.35cm}
(\mu_0)U^{^{-1}}
\hspace*{-0.35cm} (\mu,\mu_0))_{_{33}}$ (where $U(\mu,\mu_0)$ is the
renormalization group evolution between $\mu_0$ and $\mu$ at NLO) is
shown with $\mu_0=1.5\;\gev$, before and after the subtraction. It is 
clear that there is a window for $2.4\,\gev \ge \mu \ge 1.4\,\gev$ where $Z$ 
follows the expected renormalization group evolution.  
\vspace*{-0.2cm}    
\section{Very briefly on $\Delta I=1/2$ ME's}
\vspace*{-0.2cm}    
$\Delta I\!\!=\!\!1/2$ OP's mix, through penguin contractions,
also with lower dimensional OP's with
power divergent coefficients. In this case a non-perturbative
subtraction is needed (in both strategies: $K\ra\pi\pi$ or $K\ra\pi$). 
Here we will present very preliminar results for $Q^-$
with the purpose of showing that, for the first time, a 
signal has been observed (further details can be found in Ref.~\cite{guido}). 
$Q^-$ enters, together with $Q^+$, 
in the computation of $\textrm{Re}{\cal A}_{I=0}$. They are defined as\\ 
\vspace{-0.5cm}
\be
Q^{\pm} = \bar s \gamma_{\mu}^L  
u \, \bar u  \gamma^{\mu}_L d \pm  \bar s \gamma_{\mu}^L
d \, \bar u  \gamma^{\mu}_L u  - (u \to c)\,.
\ee
In our simulation the charm quark is propagating. 
Due to the GIM mechanism, the subtraction is thus 
implicit in the difference of penguin diagrams with an up quark and a
charm quark inside the loop. In Fig.~\ref{Om} the penguin contractions
of the bare $O^-$ are shown, in a kinematical configuration in which 
$M_K a=2M_\pi a=0.72$. It is interesting, even though only indicative,
to note that the ratio of bare OP's
$\<\pi\pi|Q^-|K\>_{I=0}/\<\pi\pi|Q^+|K\>_{I=2}\approx9$. In fact,
although affected by a huge statistical error and incomplete for the
lacking of the RC's and of part of the contractions, 
this result shows that penguin contractions are of the
right order of magnitude needed to explain the $\Delta I\!\!=\!\!1/2$
rule (remember that the ratio of Wilson coefficients 
$|C^-(\mu=2 \gev)/C^+(\mu=2 \gev)|\approx2$). Concerning $O^+$, 
penguin contractions gives a value close to zero (but again with large
errors). In order to reduce the statistical errors, a study to find the best 
source for the two pion state is
presently under way. 
\begin{figure}[!t]
\vspace*{-0.8cm}
\hspace*{1.4cm}
\epsfig{figure=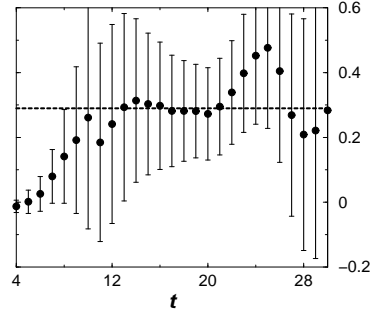,angle=-90,width=0.62\linewidth}
\vspace{-1.0cm}
\caption{\sl \small $\<\pi\pi|Q^-|K\>$ bare on 340 configurations}
\label{Om}
\vspace*{-0.7cm}
\end{figure}


}  
\end{document}